\pdfoutput=1
\documentclass[conference]{IEEEtran}

\usepackage{graphicx}
\usepackage{url}
\usepackage{algorithm, algpseudocode}
\usepackage{here}
\usepackage{amsmath}
\usepackage{acronym}
\usepackage[capitalise]{cleveref}
\usepackage{cite}

\ifCLASSINFOpdf

\else

\fi

\begin{document}
\title{High Convergence Rates of CMOS Invertible Logic Circuits Based on Many-Body Hamiltonians}

%

\author{\IEEEauthorblockN{
Naoya Onizawa\IEEEauthorrefmark{1},
Takahiro Hanyu\IEEEauthorrefmark{1}}
\IEEEauthorblockA{\IEEEauthorrefmark{1} Research Institute of Electrical Communication, Tohoku University, Sendai, Japan}
}


\maketitle

\begin{abstract}
This paper introduces CMOS invertible-logic (CIL) circuits based on many-body Hamiltonians. 
CIL can realize probabilistic forward and backward operations of a function by annealing a corresponding Hamiltonian using stochastic computing.
We have created a Hamiltonian that includes three-body interaction of spins (probabilistic nodes).
It provides some degrees of freedom to design a simpler landscape of Hamiltonian (energy) than that of the conventional two-body Hamiltonian.
The simpler landscape makes it easier to reach the global minimum energy. 
The proposed three-body CIL circuits are designed and evaluated with the conventional two-body CIL circuits, resulting in few-times higher convergence rates with negligible area overhead on FPGA.

\end{abstract}


\renewcommand\IEEEkeywordsname{keywords}
\begin{IEEEkeywords}
Stochastic computing, bidirectional operations, Ising model, simulated annealing
\end{IEEEkeywords}

%

\section{Introduction}

Invertible logic has been recently presented for providing a capability of forward and backward operations \cite{IL} as opposed to typical binary logic for only the forward operation.
The bidirectional computing capability is realized by reducing an energy (Hamiltonian) to the global minimum using random signals, where the Hamiltonian represents a function (e.g. a multiplier could be used as a factorizer in the backward mode).
Thanks to the unique feature, several challenging problems could be quickly solved, such as integer factorization (e.g. cryptography problems \cite{factorization}) and machine learning (e.g. training neural networks \cite{CIL_training,CIL_training2}).
The Hamiltonian is constructed by a network of spins (probabilistic nodes)  with interactions among them.
In the conventional implementations \cite{IL,IL_FPGA,CIL}, the interactions of the Hamiltonian are limited to the two-body one between spins.
However, the two-body Hamiltonians restrict a degree of freedom of representing a function, which could result in difficulty of reaching the global minimum energy.
Until now, the limitation of the two-body Hamiltonians and an  expansion to many-body Hamiltonians have not been studied for invertible logic.

In this paper, invertible logic circuits based on many-body Hamiltonians is introduced and compared with the conventional invertible logic based on the two-body Hamiltonians.
The proposed three-body Hamiltonians for invertible logic are obtained by linear programming, which provides simpler landscapes of Hamiltonians than the two-body one, thanks to more degree of freedom of the Hamiltonian representation.
The three-body interaction circuits are designed based on CMOS invertible logic (CIL) \cite{CIL} with stochastic computing \cite{stochastic_first,stochastic}, where the three-body interactions require multiplication among spin states.
As stochastic computing realizes multiplication using a logic gate, the three-body invertible logic circuits can be simply implemented.
Our contribution of the proposed three-body invertible logic over the conventional two-body invertible logic  is: 1) few-times higher convergence rates in several invertible logic circuits and 2) negligible  area overhead on FPGA.

\section{Invertible Logic Based on Many-Body Hamiltonian}

\subsection{Invertible logic based on two-body Hamiltonian}

\begin{figure}[t]
	\centering
	\includegraphics[width=0.7\linewidth]{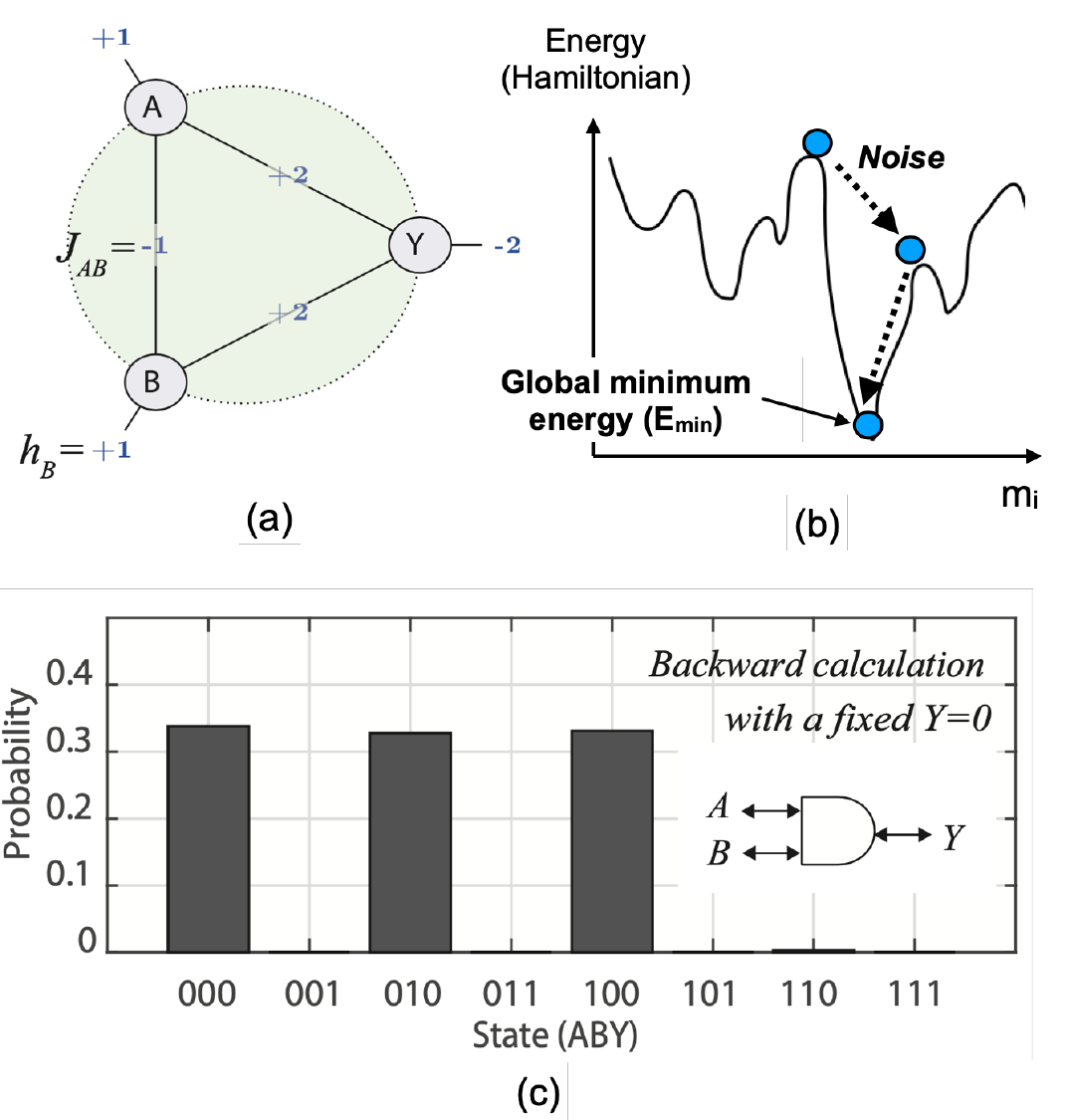}
		\vspace{-3mm}
	\caption{Two-input invertible AND in one example of invertible logic: (a) two-body Hamiltonian, (b) landscape of energy (Hamiltonian), and (c) state probabilities when $Y$ is fixed to 0 at backward mode.}
	\label{fig:IL}
	\vspace{-1mm}
\end{figure}

Invertible logic circuits operate at forward and/or backward modes between inputs and outputs using Boltzmann machine \cite{Boltzmann1984}  and probabilistic bits (p-bits) \cite{IL}, which are also called spins.
%
%
A function for the forward/operation modes is embedded using a Hamiltonian including a spin bias ($h$) and weights between spins ($J$).
%
%
A two-body Hamiltonian of invertible logic is given by:
\begin{equation}
	H = -\sum_i h_i m_i - \sum_{i<j} J_{ij} m_i m_j.
	\label{eqn:H_2body}
\end{equation}
where a spin state is  $m_i \in \{-1,1\} \ (1 \leq i \leq l$) and $l$ is the number of spins.
Differing from invertible logic is reversible logic that are constructed of special gates (such as Controlled NOT (CNOT) or Toffoli gates) having a direct one-to-one mapping of inputs to outputs \cite{RL_survey}.
While  both reversible and invertible logic reconstruct inputs from a given output value, they differ at fundamental levels (see details in \cite{IL_design_framework}).

\cref{fig:IL} (a) shows a two-body Hamiltonian of a two-input invertible AND ($Y=A\cap B$).
The Hamiltonian is obtained based on ground state-spin logic \cite{GSS1,GSS2}.
It is an energy, where valid states of a function are embedded to the global minimum energy ($E_{min}$) and invalid sates are embedded to local minimum energies.
With given $h$ and $J$, each spin updates $m_i$ using a random signal to reach $E_{min}$, as illustrated in \cref{fig:IL} (b).
$m_i$ is given by:
\begin{subequations}
	\begin{equation}
	m_i\left(t+\tau\right) \simeq \rm{sgn}\Big(\rm{tanh}\big( I_i\left(t+\tau\right) \cdot I_0 \big)\Big),
	\label{eqn:SP_2body1}
	\end{equation}
	\begin{equation}
	I_i\left(t+\tau\right) \simeq \Big(h_i+\sum_j J_{ij}m_j\left(t\right) + w_{rnd} \cdot \rm{sgn}\Big(rnd\big(-1,+1\big)\Big)\Big),
	\label{eqn:SP_2body2}
	\end{equation}
	\label{eqn:SP_2body}
\end{subequations}
where $I_0$ is a pseudo inverse temperature to control invertible logic and $w_{rnd}$ is a weighted value for random signals.
\cref{fig:IL} (c) shows an example of the two-input invertible AND in the backward mode.
With fixing the output ($Y$) to `0' ($``m_y = -1"$), there are three valid states (`ABY') of (`000', `010', `100').
In this simulation, the three valid states appear with almost the same probability of 33\%.

\subsection{Design of many-body Hamiltonians}
	
	\begin{figure}[t]
		\centering
		\includegraphics[width=0.7\linewidth]{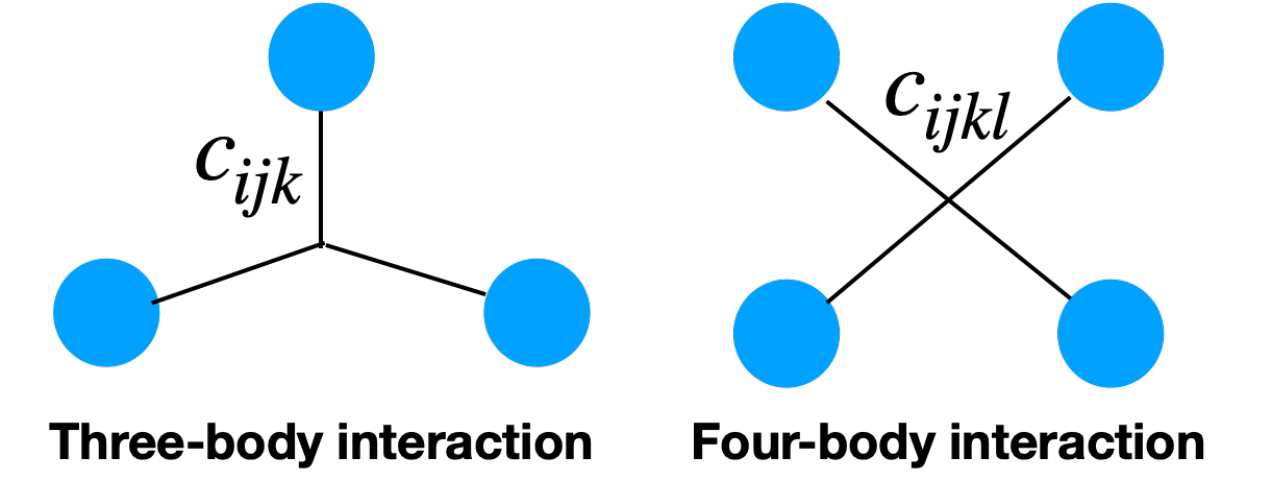}
			\vspace{-3mm}
		\caption{Many-body Hamiltonians with interactions among more than two spins.}
		\label{fig:many_body}
		\vspace{-1mm}
	\end{figure}
	
There exists a limitation of representing a function of invertible logic due to the two-body Hamiltonian, resulting in difficulty of reaching the global minimum energy.
In this paper, invertible logic based on many-body Hamiltonians is introduced to provide more degree of freedom of representing Hamiltonians for fast convergence of invertible logic.
\cref{fig:many_body} illustrates interactions of three-body and four-body Hamiltonians.
Unlike the two-body Hamiltonian, there are interactions among more than two spins.
The many-body Hamiltonian is given by:
	\begin{equation}
	H = -\sum_i c_i m_i - \sum_{i<j} c_{ij} m_i m_j - \sum_{i<j<k} c_{ijk} m_i m_j m_k \cdots .
	\label{eqn:H_many_body}
	\end{equation}
where $c_i$, $c_{ij}$, and $c_{ijk}$ are interaction coefficients of one-body, two-body and three-body Hamiltonians, respectively.

\begin{table}[t]
	\caption{Proposed three-body Hamiltonian of two-input invertible AND ($Y=A\cap B$).}
	\label{tb:3bodyH_IAND}
	\centering
	\begin{tabular}{c|c|c|c|c|c|c}
		\hline
		$c_A$ & $c_B$ & $c_Y$ & $c_{AB}$ & $c_{AY}$ & $c_{BY}$ & $c_{ABY}$ \\
		\hline
		0 & 0 & -1 &  0 & 1 & 1 & 1 \\
		\hline
	\end{tabular}
	\vspace{0mm}
\end{table}

The Hamiltonian coefficients are obtained based on linear programming that is an extended method for the two-body Hamiltonian \cite{IL_design_framework}.
	In invertible logic for a function with inputs ($x_i \in \{0,1\} \ (1 \leq i \leq p))$ and outputs ($y_i \in \{0,1\} \ (1 \leq i \leq q))$, the energies (Hamiltonians) of the valid states of the function must be equal to the global minimum ($E_{min}$) while that of the invalid states are larger than $E_{min}$.
	For the three-body Hamiltonian, the energies of the valid and the invalid states are given as follows:
	\begin{eqnarray}
	E_k = 
	\begin{cases}
	- \sum_i c_i m_i - \sum_{i < j} c_{ij} m_i m_j \\  
	\hspace{5mm} - \sum_{i < j < k} c_{ijk} m_i m_j m_k = E_{min} \\ 
	\hspace{15mm} (f(x_1...x_p)=(y_1...y_q)) \\
	- \sum_i c_i m_i - \sum_{i < j < k} c_{ij} m_i m_j \\ 
	\hspace{5mm} - \sum_{i < j < k} c_{ijk} m_i m_j m_k  \geq E_{min} + d \\
	\hspace{15mm} (otherwise)
	\end{cases}
	\label{eqn:LP1}
	\end{eqnarray}
	where $d$ is the energy difference between $E_{min}$ and the second minimum energy.
	The equation can be extended to more than three-body Hamiltonians.
	The objective function is maximizing $d$ using linear programming in order to obtain $c_i$, $c_{ij}$ and $c_{ijk}$ as follows:
	\begin{eqnarray}
	\rm{maximize} & d \\
	\rm{subject \ to} & Eq. (\ref{eqn:LP1}),
	\label{eqn:LP2}
	\end{eqnarray}
	where $m_i$, $m_j$, and $m_k$ are  constants and $c_i$, $c_{ij}$, ,$c_{ijk}$, $E$ and $d$ are variables.
	\cref{tb:3bodyH_IAND} summarizes the three-body Hamiltonian coefficients of the two-input invertible AND based on linear programming.

	\begin{figure}[t]
	\centering
	\includegraphics[width=0.8\linewidth]{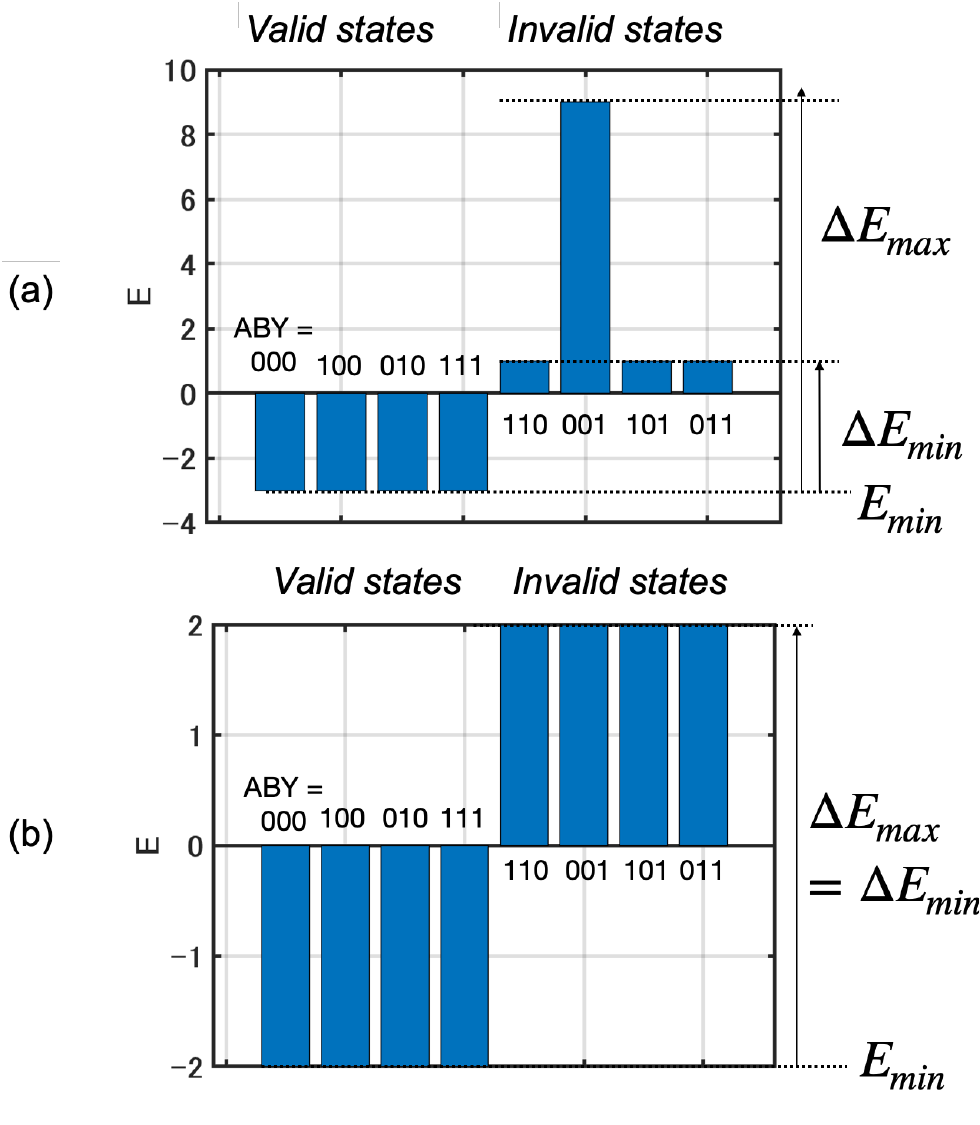}
	\caption{Energy of two-input invertible AND ($Y=A\cap B$): (a) two-body Hamiltonian and (b) three-body Hamiltonian. The three-body Hamiltonian provides a simpler energy landscape than the two-body one.}
	\label{fig:energy}
	\vspace{0mm}
	\end{figure}

\begin{table}[t]
	\caption{Energy of invertible logic gates based on two-body and three-body Hamiltonians.}
	\label{tb:energy}
	\centering
	\begin{tabular}{c|c||c|c|c|c}
		\hline
		\multicolumn{2}{c||}{} &  $E_{min}$ & $\Delta E_{min}$ & $\Delta E_{max}$ & $N_{EL}$ \\
		\hline
		\hline
		Two-body& AND, OR & -3 & 4 & 12 & 3\\
		& XOR, XNOR & -4 & 2 & 18 & 4 \\
		\hline
		Three-body & AND, OR  & -2 & 4 & 4 & 2 \\
		& XOR, XNOR & -2 & 4 & 4 & 2 \\
	    \hline
	\end{tabular}
	\vspace{-2mm}
\end{table}

\cref{fig:energy} compares the energies of the two-input invertible AND based on the two-body and the three-body Hamiltonians.
In the two-body invertible AND, $E_{min}$ is `-3' for the valid states while energies of the invalid states are `1' and `9'.
There are three energy levels ($N_{EL}$) in total.
Note that $\Delta E_{min}$ is an energy difference between $E_{min}$ and the second minimum energy, and $\Delta E_{max}$ is an energy difference between $E_{min}$ and the maximum energy.
Due to the limitation of representing a function in the two-body Hamiltonian, a complicated energy landscape could be obtained.
In contrast, the energy landscape of the three-body invertible AND is simpler than that of the two-body one because of more degree of freedom of representing the Hamiltonian, which contains only two energy levels of `-2' (valid sates) and `2' (invalid states) with $N_{EL} = 2$.
\cref{tb:energy} compares the energies of invertible logic circuits based on the two-body and the three-body Hamiltonians.
The three-body Hamiltonians can provide simpler energy landscapes, which could result in higher convergence of invertible logic.

\section{Circuit Implementation of three-body CMOS invertible-logic (CIL) circuits}

An invertible logic circuit for the three-body Hamiltonian is introduced in this section.
Invertible logic circuits for the two-body Hamiltonian have been designed using a probabilistic model \cite{IL}, binary logic \cite{IL_FPGA}, or stochastic computing \cite{CIL}.
The invertible logic circuits for the three-body Hamiltonian are designed based on stochastic computing in this paper.
With given $c_i$, $c_{ij}$ and $c_{ijk}$, each spin updates $m_i$  to reach $E_{min}$ as follows:
\begin{subequations}
	\begin{equation}
	m_i\left(t+\tau\right) \simeq \rm{sgn}\Big(\rm{tanh}\big( I_i\left(t+\tau\right) \cdot I_0 \big)\Big),
	\label{eqn:SP_3body1}
	\end{equation}
	\begin{eqnarray}
	I_i\left(t+\tau\right) &\simeq& \Big(c_i+\sum_j c_{ij}m_j\left(t\right)  + \sum_{j,k} c_{ijk}m_j\left(t\right)m_k\left(t\right)\nonumber \\
	&&+ w_{rnd} \cdot \rm{sgn}\Big(rnd\big(-1,+1\big)\Big)\Big).
	\label{eqn:SP_3body2}
	\end{eqnarray}
	\label{eqn:SP_3body}
\end{subequations}

	\begin{figure}[t]
	\centering
	\includegraphics[width=0.8\linewidth]{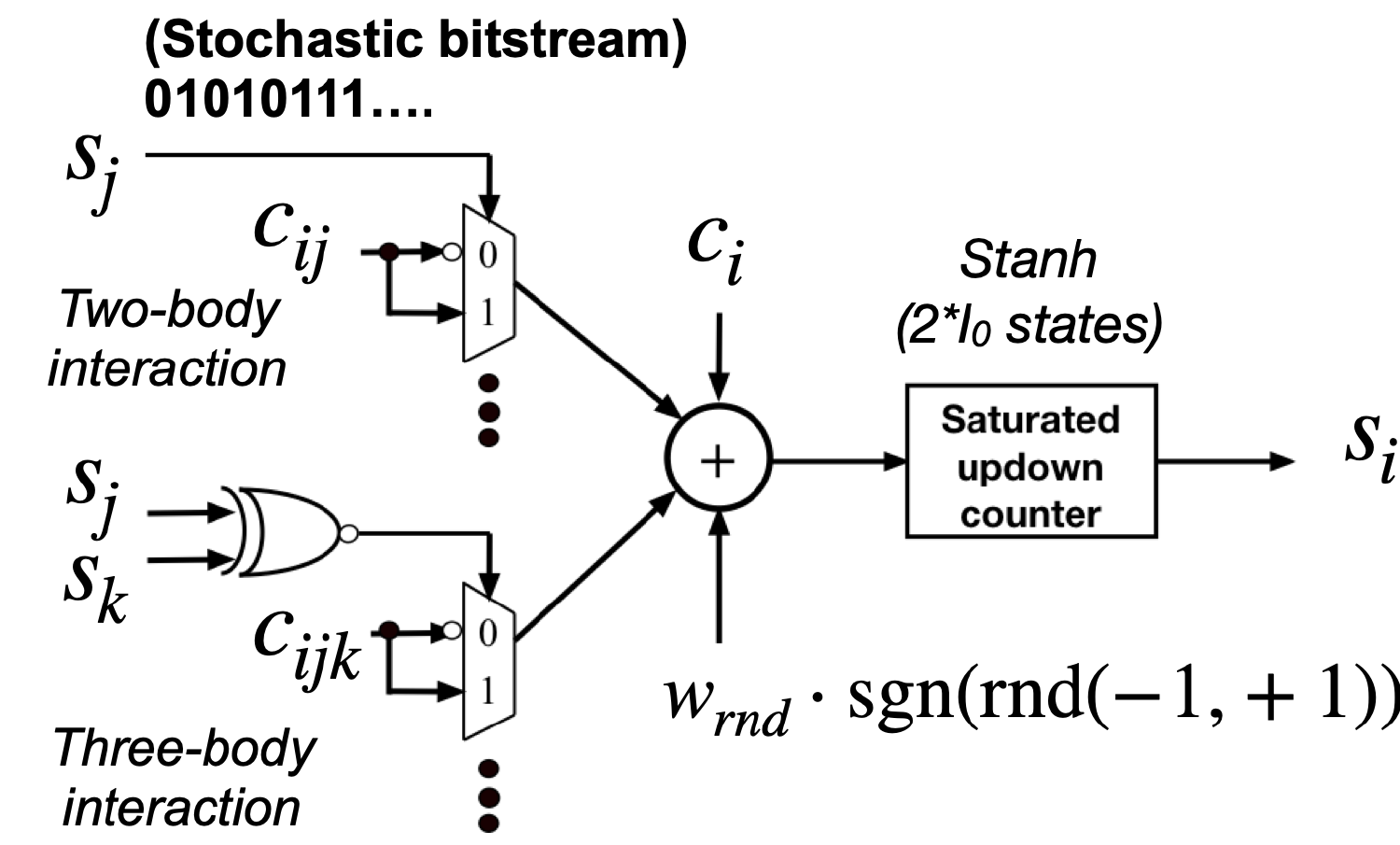}
		\vspace{-3mm}
	\caption{Proposed spin-gate circuit for three-body Hamiltonian based on integral stochastic computing that corresponds to \cref{eqn:SP_3body} with $s_i = \frac{1+m_i}{2}$.}
	\label{fig:circuit}
	\vspace{-1mm}
\end{figure}

\cref{fig:circuit} describes a spin-gate circuit for the three-body Hamiltonian based on stochastic computing  that corresponds to \cref{eqn:SP_3body}.
Stochastic computing uses values represented by frequencies of `1' in bit streams \cite{stochastic_first,stochastic} and has been used for several applications, such as low-density parity-check decoders, image processing, digital filters, and deep neural networks \cite{SLDPC,Simage,SIIR,SDNN}.
Let us denote by $S \ \in \{0,1\}$ a random bit streams.
A real number, $s \in [-1:1]$, is represented by ($2*\mathrm{E}[S]-1$) in binary stochastic computing in bipolar format, where $\mathrm{E}[S]$ denotes the expected value of the random variable, $S$.
The spin state, $m_i$, corresponds to $(2s_i-1)$.

Addition, multiplication, and tanh of \cref{eqn:SP_3body} are designed based on integral stochastic computing \cite{SDNN}.
In integral stochastic computing, one or more bit streams are concurrently used to represent data values in larger ranges than that of binary stochastic computing.
Let us denote by $X \ \in \{-r,-(r-1), ... ,r\}$ a random bit streams ($r \in \{1, 2, ...\}$).
A real number, $x \in [-r:r]$, is represented by $\mathrm{E}[X]$ in signed format, where $\mathrm{E}[X]$ denotes the expected value of the random variable, $X$.
The multiplication for the two-body interactions ($m_jc_{ij}$)  is designed using a two-input multiplexor.
In addition, the multiplication for the three-body interactions ($m_jm_kc_{ijk}$)  is simply designed using a two-input multiplexor and an XNOR gate, which realizes multiplication of two stochastic bit streams ($m_jm_k$).
The tanh function is approximated using a saturated updown counter.
The random signals can be generated using several methods, such as linear feedback shift register and xorshift \cite{CIL}.
Due to the limited space of the short paper, please see the details of stochastic computing in \cite{stochastic,stochastic_book}.

\section{Evaluation}

\subsection{Convergence rate to the global minimum energy}

	\begin{figure}[t]
	\centering
	\includegraphics[width=0.7\linewidth]{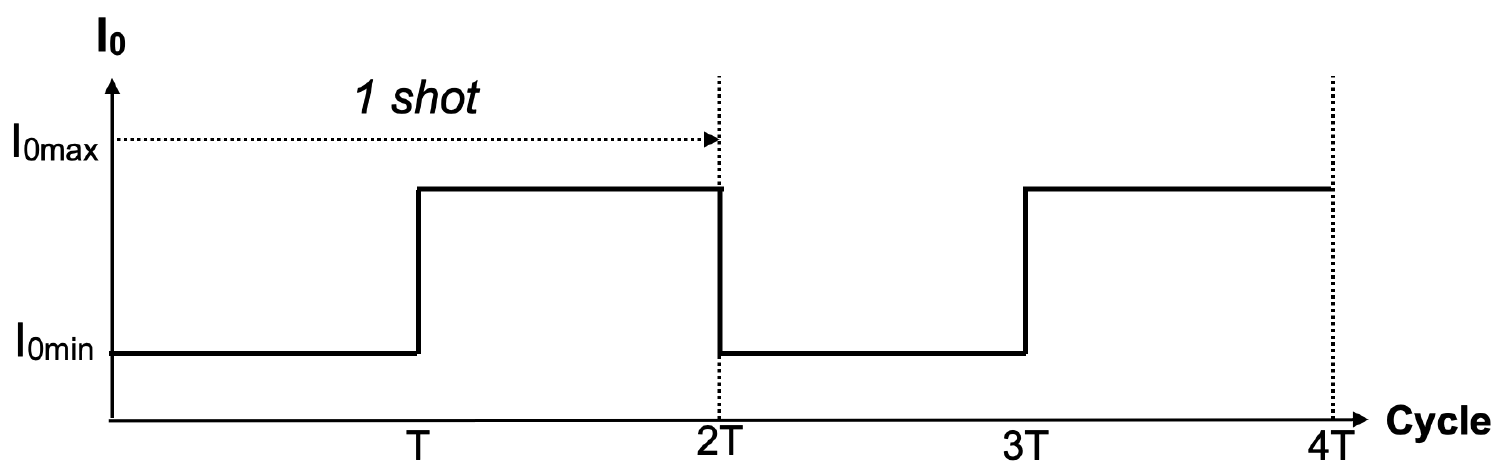}
    \vspace{-3mm}
	\caption{$I_0$ control for simulating invertible adders, where $I_{0min}$=2, $I_{omax}$=4, and $T$=100 are selected in this paper.}
	\label{fig:I0}
	\vspace{-1mm}
\end{figure}

In order to evaluate the effects of many-body Hamiltonian on invertible logic, invertible adders ($Y = A + B$) are designed on the two-body and the three-body Hamiltonians.
An architecture of the invertible adders derives from a ripple carry adder designed using logic gates.
A Hamiltonian of each logic gate is used for the two-body interactions in \cite{CIL} and the three-body interactions summarized in \cref{tb:energy}.
The invertible adders are simulated using MATLAB and the convergence rates to $E_{min}$ are evaluated with 100 trials at the backward mode of fixing $Y$.
$I_0$ (pseudo inverse temperature) is controlled shown in \cref{fig:I0} with a shot of a pulse signal of $I_{0max}$ and $I_{0min}$ during $2T$ cycles.
$I_{0min}$ can induce more variations of spin states and $I_{0max}$ can stabilize the spin sates to converge to $E_{min}$.
The shot is iteratively applied to the invertible adders with the maximum number of shots, $N_{shot}$.
In this paper, $I_{0min}=2$, $I_{0max}=4$, and $T=100$ are selected, where $w_{rnd}$ is 3 and  $\tau$ is 1 in \cref{eqn:SP_3body}.

	\begin{figure}[t]
	\centering
	\includegraphics[width=1.0\linewidth]{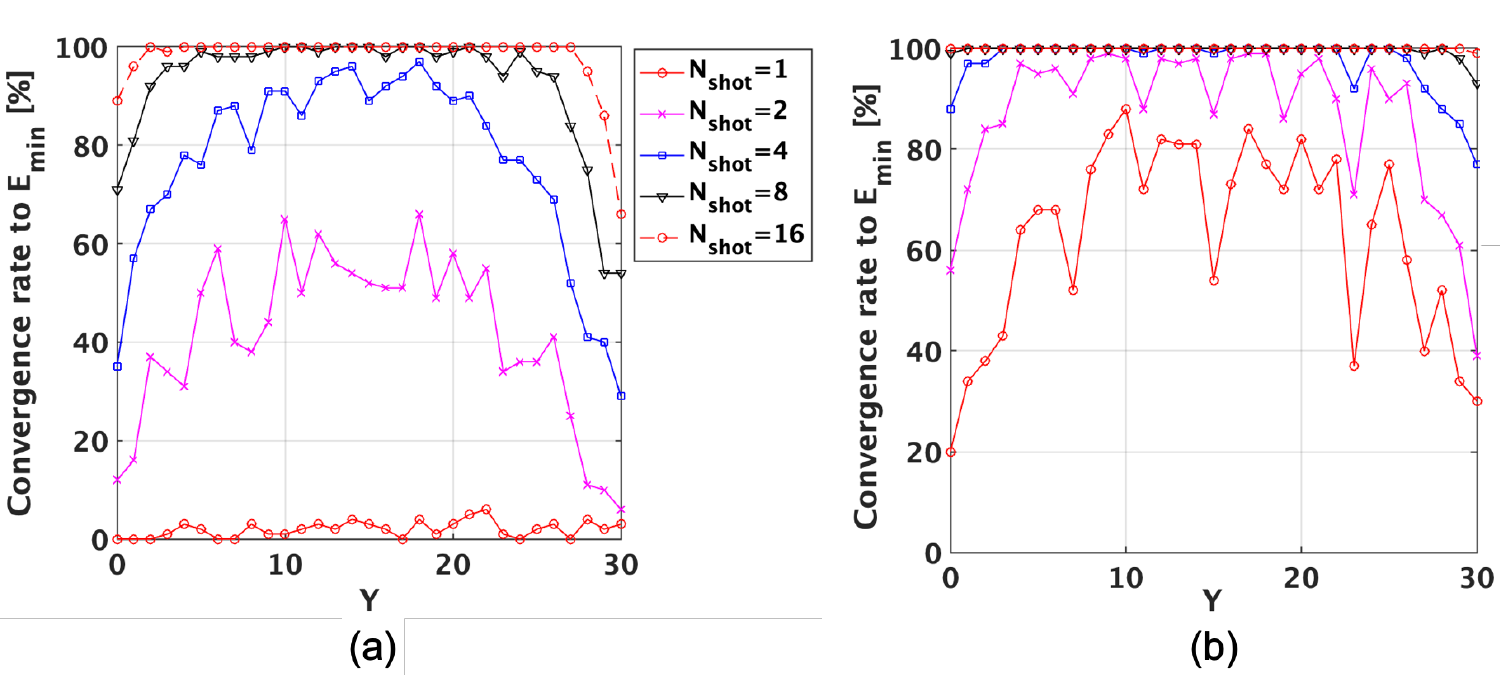}
	\vspace{-5mm}
	\caption{Convergence rates of 4-bit invertible adder ($Y = A + B$) at backward mode with a fixed output, $Y$: (a) two-body and (b) three-body Hamiltonians. The three-body invertible logic realizes few-time higher convergence rates than the two-body one.}
	\label{fig:conv_4b_adder}
	\vspace{0mm}
\end{figure}


\cref{fig:conv_4b_adder} shows the convergence rates of 4-bit invertible adder at backward mode based on the two-body and the three-body Hamiltonians.
For the backward operation, the output, $Y$, is fixed to obtain inputs with the global minimum energy, $E_{min}$.
For the two-body 4-bit invertible adder, $N_{shot}$ of 16 is required to reach $E_{min}$ at almost all the cases.
In contrast, the three-body 4-bit invertible adder reaches $E_{min}$ with $N_{shot}$ of 4 at almost all the  cases, which can provide four-times faster convergence.
%
%
The high convergence rates derive from the simpler energy landscapes of the three-body Hamiltonians explained in \cref{fig:energy}.

	\begin{figure}[t]
	\centering
	\includegraphics[width=1.0\linewidth]{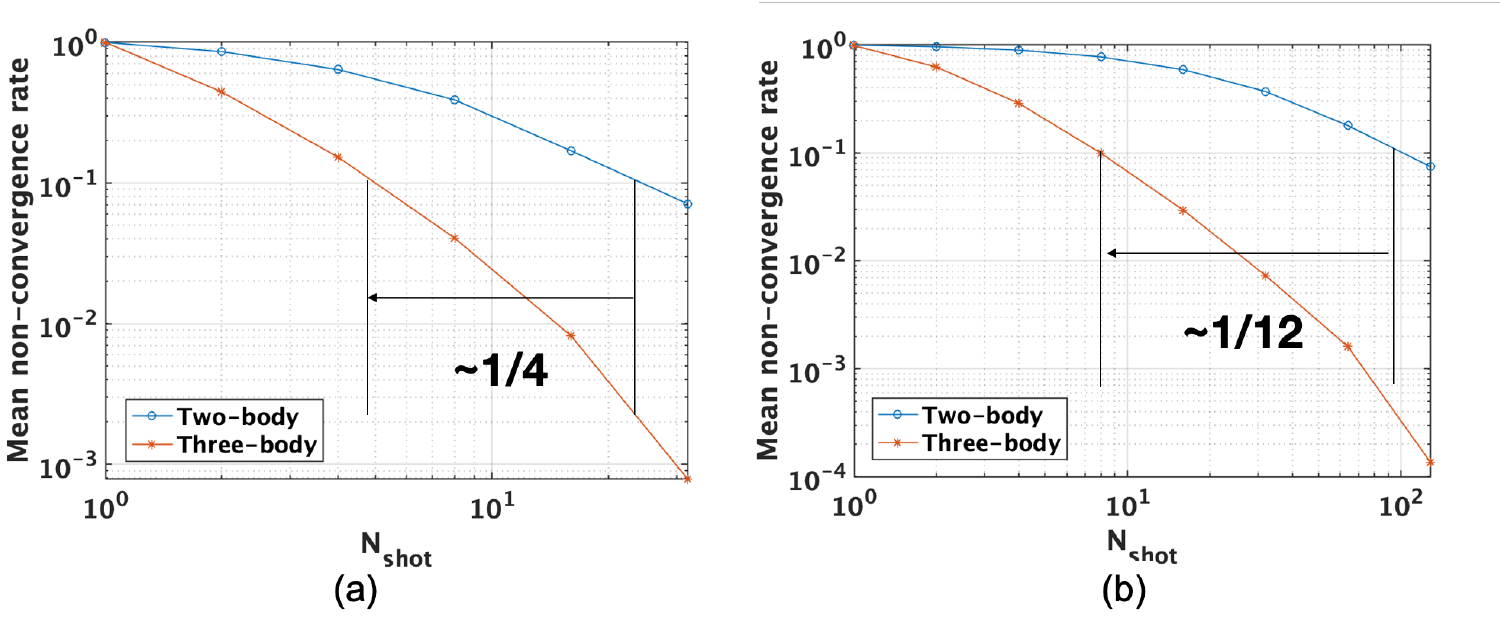}
	\vspace{-5mm}
	\caption{Mean non-convergence rates of invertible adders at backward mode: (a) 6-bit and (b) 8-bit. Faster convergence is realized by the proposed three-body Hamiltonians.}
	\label{fig:nonconv_rate}
	\vspace{0mm}
\end{figure}

\cref{fig:nonconv_rate} shows mean non-convergence rates of the invertible adders at the backward mode.
The mean non-convergence rates are calculated using the convergence rates of all the cases of $Y$.
At the mean non-convergence rate of $10^{-1}$, the three-body invertible adders achieve the reduction of $N_{shot}$ to roughly 1/4 and 1/12 in the 6-bit and the 8-bit invertible adders, respectively in comparison with the two-body invertible adders.
%

\subsection{Hardware results}

\begin{table}[t]
	\caption{Hardware evaluation of spin-gate circuits at 100 MHz on Xilinx Kintex-7 FPGA (XC7K325T-2FFG900C).}
	\label{tb:performance}
	\centering
	\begin{tabular}{c||c|c||c|c}
		\hline
		&  \multicolumn{2}{c||}{Conventional} &  \multicolumn{2}{c}{Proposed} \\
		&  \multicolumn{2}{c||}{(two-body)  \cite{CIL}} &  \multicolumn{2}{c}{(three-body)} \\
		\cline{2-5}
		Number of inputs& LUT & FF & LUT & FF \\
		\hline
	2&74&	11&	78&	11
\\
	6&105&	11&	122&	11
\\
	10&144&	11&	163&	11 \\
		\hline
	\end{tabular}
	\vspace{0mm}
\end{table}

\cref{tb:performance} summarizes the hardware cost of the two-body and the three-body spin-gate circuits on FPGA.
Both circuits are designed using SystemVerilog and synthesized using Xilinx VIvado 2018.2 for Kintex-7 FPGA (XC7K325T-2FFG900C) with 100 MHz.
The three-body spin-gate circuit requires roughly 10\% more look-up tables (LUTs) than the two-body spin-gate circuit.
The hardware overhead comes from the extra XNOR gate illustrated in \cref{fig:circuit} while the other circuit blocks are identical between the two-body and the three-body invertible logic.
When the number of inputs increases, the ratio of the hardware overhead is almost equivalent.
As a result, it is expected that the proposed three-body invertible logic circuits could be more effective in larger Hamiltonians for fast convergence with negligible hardware overhead.

\section{Conclusion}

The three-body invertible logic has been presented for the high convergence rates in comparison with the conventional two-body invertible logic.
The three-body Hamiltonians provide more degree of freedom of representing bidirectional functions, resulting in the simpler energy landscapes and few-time higher convergence rates in the invertible adders.
The area overhead due to the three-body interactions is negligibly small (e.g. 10\% on FPGA), thanks to the area-efficient multiplication based on stochastic computing.
n the future prospect, it is important to evaluate the three-body invertible logic circuits for large applications, such as integer factorization and machine learning.

\section*{Acknowledgment}

This work was supported by JST PRESTO Grant Number JPMJPR18M5 and CANON MEDICAL SYSTEMS CORPORATION.

\bibliographystyle{IEEEtran}
\bibliography{ISCAS2021}

\end{document}